\documentclass{iauc}
\usepackage{graphicx}

\title[Ground-based differential imager] 
{Analysis of ground-based differential imager performance}

\author[Boccaletti et al.]   
{A. Boccaletti$^1$, D. Mouillet$^2$, T. Fusco$^3$, P. Baudoz$^1$ \break 
C. Cavarroc$^1$, J.-L. Beuzit$^4$,  C. Moutou$^5$ \and K. Dohlen$^5$}

\affiliation{$^1$LESIA, Observatoire de Meudon, 5 pl. J. Janssen, 92195 Meudon, France \break 
email: anthony.boccaletti@obspm.fr \\[\affilskip]
$^2$Observatoire Midi-Pyr\'en\'ees, 57 av. d'Azereix, 65008 Tarbes, France \break 
$^3$ONERA, 29 avenue de la Division Leclerc, 92320 Chatillon, France\break
$^4$LAOG, Observatoire de Grenoble, 38041 Grenoble, France\break
$^5$LAM, Traverse du Siphon, 13376 Marseille, France\break
}

\pubyear{2005}
\volume{xxx} 
\pagerange{119--126}
\date{?? and in revised form ??}
\jname{Proceedings Title IAU Colloquium }
\editors{A.C. Editor, B.D. Editor and C.E. Editor, eds.}
\begin{document}

\maketitle

\begin{abstract}
In the context of extrasolar planet direct detection, we evaluated the performance of differential imaging with ground-based telescopes. This study was carried out in the framework of the VLT-Planet Finder project and is further extended to the case of Extremely Large Telescopes. Our analysis is providing critical specifications for future instruments mostly in terms of phase aberrations but also regarding alignments of the instrument optics or offset pointing on the coronagraph. It is found that Planet Finder projects on 8m class telescopes can be successful at detecting Extrasolar Giant Planets providing phase aberrations, alignments and pointing are accurately controlled. The situation is more pessimistic for the detection of terrestrial planets with Extremely Large Telescopes for which phase aberrations must be lowered at a very challenging level.

\end{abstract}

\firstsection 
\section{Future ground-based instruments to search for exoplanets}
\begin{figure}[t]
   \centering
      \includegraphics[width=8cm]{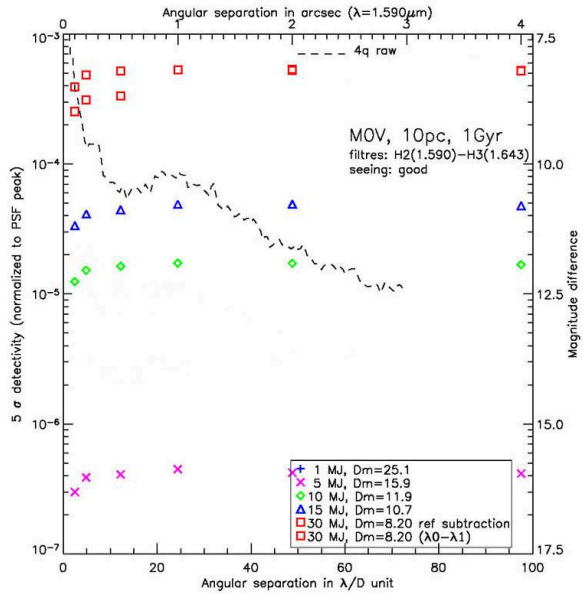}
   \caption{Raw contrast achieved with an 8m telescope, a  40$\times$40 actuators AO system and an achromatic phase mask coronagraph. The expected intensities of planets are over-plotted for several masses and several angular separations. The contrast at 0.5" is only $6.10^{-5}$ which is not allowing the detection of planetary masses. A speckle pattern calibration unit is definitely needed to improve the contrast.}
   \label{fig:rawcontrast}
\end{figure}

Many techniques and projects were proposed in the past few years to image directly extrasolar planets from the ground. Specific instruments like Adaptive Optics (AO) systems and coronagraphs were developed in this way. However, it is expected that even an extreme AO system coupled with a high rejection coronagraph will not be sufficient to attenuate the diffracted light form the on-axis star down to the level of gazeous planets like Jupiter. For illustration, Fig. \ref{fig:rawcontrast} shows the 5$\sigma$ detectivity radial profile considering a 40$\times$40 actuators AO system on an 8m telescope and an achromatic phase mask coronagraph. The symbols on the plot correspond to different planet masses assuming an old M0V star at 10pc observed in the H band. The contrast at 0.5" (corresponding to a 5AU orbit at 10pc) is only $6.10^{-5}$. In that case, the detection of planets (with a mass lower than 13M$_J$) is definitely precluded. The situation would be more favorable for younger stars but not sufficient to explore the very low masses at close angular separations. 

The radial contrast presented in Fig. \ref{fig:rawcontrast} arises from a residual diffraction pattern 
made with atmospheric speckles left uncorrected by the AO system plus static speckles mostly originating from the non common path. It is worth to remind here that a coronagraph has the ability to suppress the coherent light only and then random interferences like speckles are escaping its action.
Nevertheless, several smart techniques of speckle calibrations were already proposed. The most developed so far is probably the spectral differential imaging (\cite{marois00}) first implemented inside TRIDENT at CFHT  (\cite{marois05}) and more recently at the VLT (\cite{lenzen05}). The basic idea is to record simultaneously two images in two different but spectrally close filters on the same detector. If phase aberrations are small, the images are identical once rescaled in intensity and spatially matched (speckle pattern has a radial dependence with the wavelength). But they actually differs since the instrument has always chromatic aberrations and because the 2-beam separations implies 2 different optics which means differential phase aberrations. 
The question is: how much the detection level can be lowered by this technique when realistic assumptions are made on the whole system (atmosphere, telescope, instrument) ?
 
In this paper, we present some results of numerical simulations we performed to assess the performance of differential technique. Section 2 describes the principle of the simulation we performed in the case of the VLT Planet Finder project and gives the list of parameters we have considered. Section 3 shows some examples of sensitivity analysis for some particular parameters. Finally, a similar study is carried out for the particular case of Extremely Large Telescopes.
\section{ Modeling differential imaging}
The current design of the VLT Planet Finder includes an extreme AO system (40$\times$40 actuators) to achieve large Strehl ratios, several coronagraphic masks to block out the on-axis starlight, an appropriate stop to cover the residual diffraction inside the pupil and a dual band imager to calibrate the speckle pattern in real time according to the technique of Spectral Differential Imaging (Marois et al. 2000).

To be realistic the simulation must include a thorough model of the atmosphere, of the telescope, and of the instrument. The simulation is performed with a sequence of 3 Fourier transforms starting from the first pupil plane and up to the detector plane.
For the telescope, we considered the actual phase maps as provided by ESO (for M1, M2 and M3 telescope mirrors). For the instrument, we took into account the number of optical surfaces we have in our optical design on which we added a f$^{-2}$ power spectrum density (PSD) phase defects. Upstream the dichroic of the AO, the phase maps (including the atmosphere and the telescope) were filtered by the PSD of the AO system (below the cutoff frequency). Similarly, phase maps upstream the coronagraph were filtered by the PSD of the phase diversity algorithm (36 Zernike polynomials are assumed perfectly corrected). The differential optics in front of the detectors were not filtered. When relevant, we also considered the chromatism of the coronagraph (for phase masks). 

In addition, we included differential offset pointing (on the coronagraph) and differential misalignment (of the telescope pupil with respect to the instrument pupil). The term ÓdifferentialÓ refers to a chromatic and temporal variation. In other words, the pointing will be slightly different in the 2 simultaneous spectral channels but also when observing sequentially a reference star as required to calibrate the differential optics. We also accounted for possible misalignment of the Lyot stop of the coronagraph.

A second numerical code models the photon noise, the detector noise (readout and flat field) and the background noise. Images are normalized in intensity using numerical models for several ages and masses of the planets. Then, we provided several types of data: raw coronagraphic images; single subtraction after proper rescaling in size and intensity following the principle of spectral differential imaging ($I_s(\lambda{_1})-I_s(\lambda_2)$ where $I_s$ and $I_r$ denotes the image of the star and of the reference); and double subtraction obtained by combining the single subtraction and the calibration on a reference star ($[I_s(\lambda_1)-I_s(\lambda_2)]-[I_r(\lambda_1)-I_r(\lambda_2)]$). Single subtraction provides calibration of the common aberrations (both atmospheric and instrumental) but double subtraction has also the advantage to provide calibration of the differential aberrations.

\begin{figure}[t]
   \centering
      \includegraphics[width=12cm]{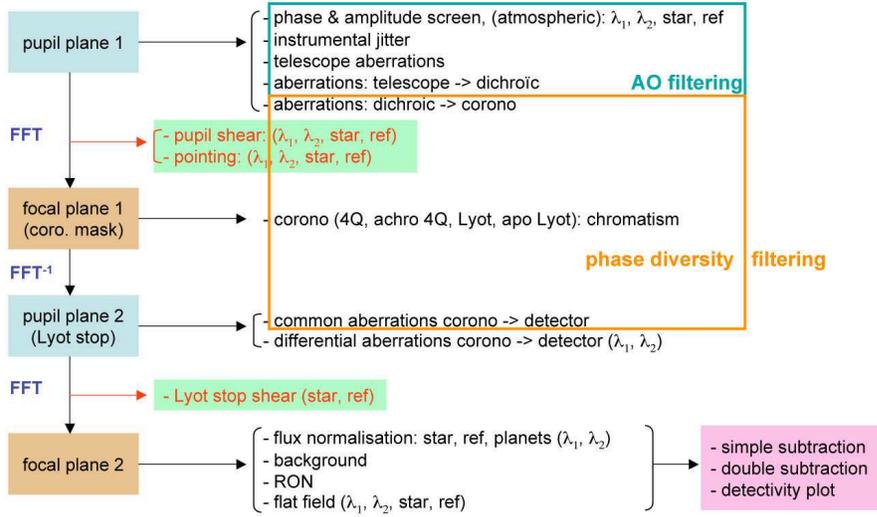}
   \caption{Flow chart of the numerical simulation presenting the main steps at the left and a more detailed structure at the right. The simulation is divided in 2 codes, one for calculating the diffraction images and the second one for including photometry. As outputs, the simulation provides detectivity plots at 5$\sigma$ as a function of the angular separation.}
   \label{fig:flowchart}
\end{figure}

The flow chart of the simulations is shown in Fig. \ref{fig:flowchart}.
\section{Sensitivity analysis}
\begin{figure}[t]
   \centering
      \includegraphics[width=6.5cm]{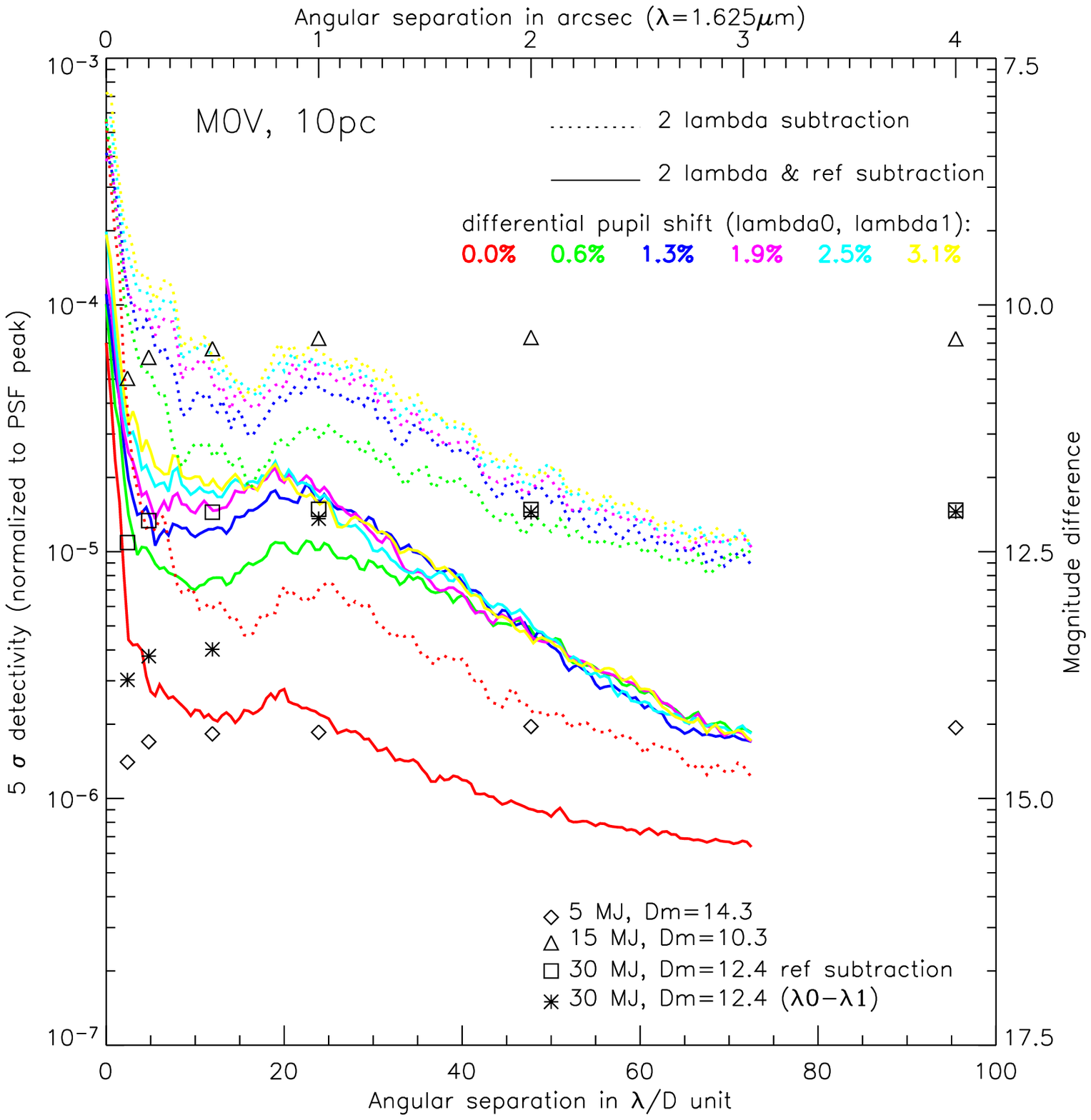}
      \includegraphics[width=6.5cm]{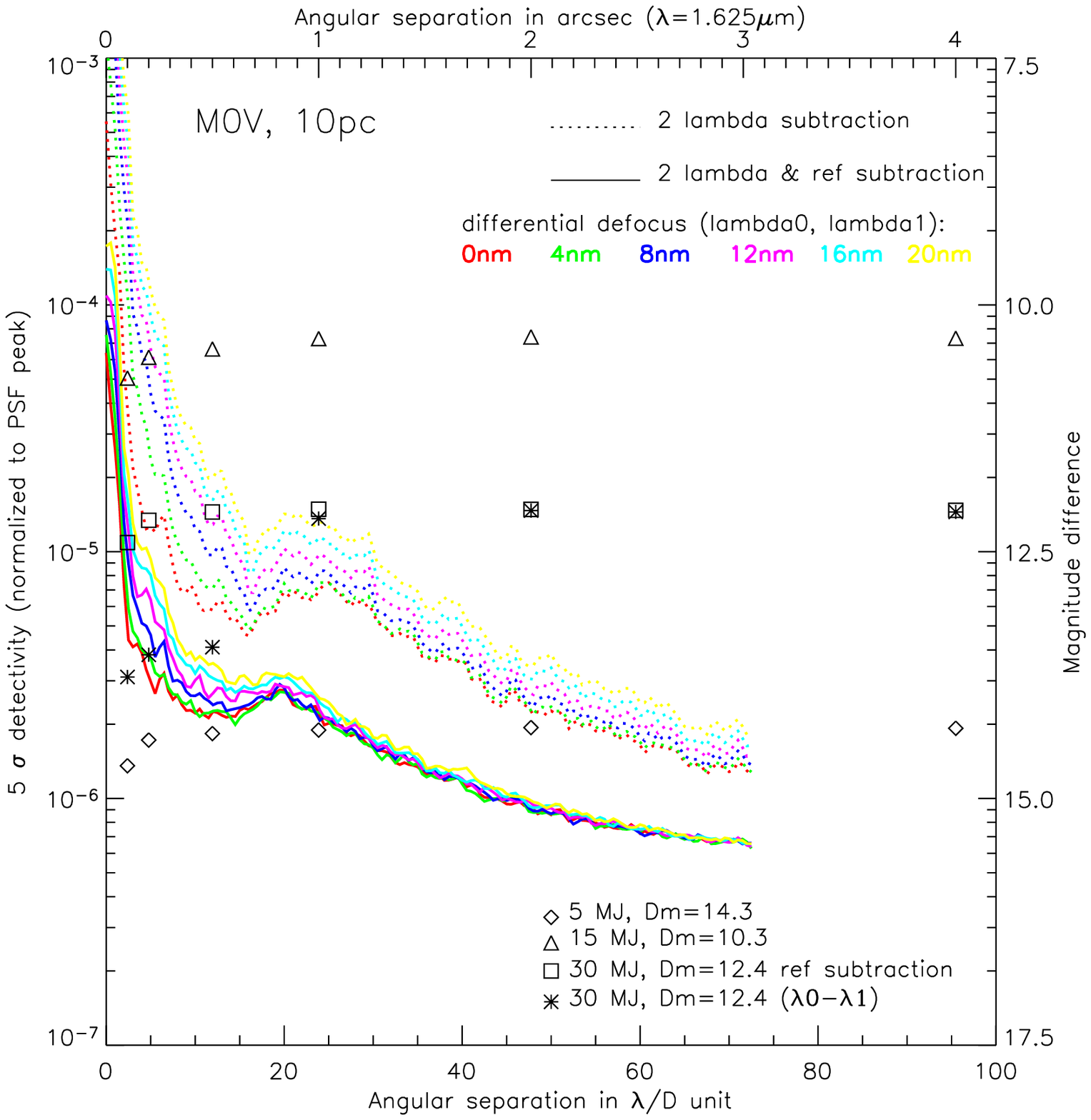}
   \caption{Sensitivity analysis of the chromatic pupil shear between the telescope and the instrument (left) and of the chromatic defocus (right). Dotted lines give the radial contrast at 5$\sigma$ for a single subtraction and solid lines are for double subtraction. The pupil shear is given in percent of the pupil diameter and the defocus is expressed in nanometers rms. Color codes are detailed on the plots. }
   \label{fig:sensitivity1}
\end{figure}

\begin{figure}[t]
   \centering
      \includegraphics[width=6.5cm]{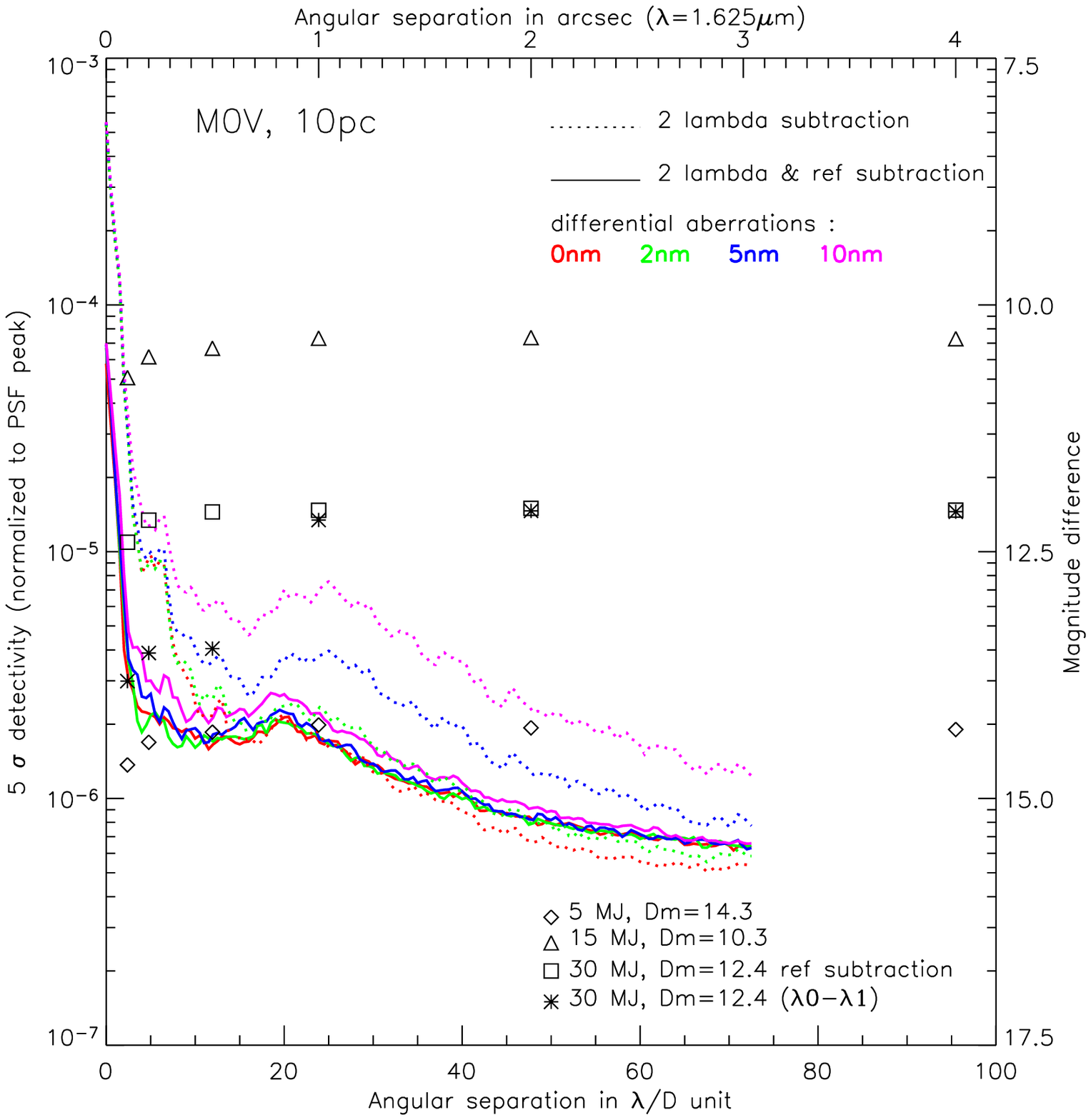}
      \includegraphics[width=6.5cm]{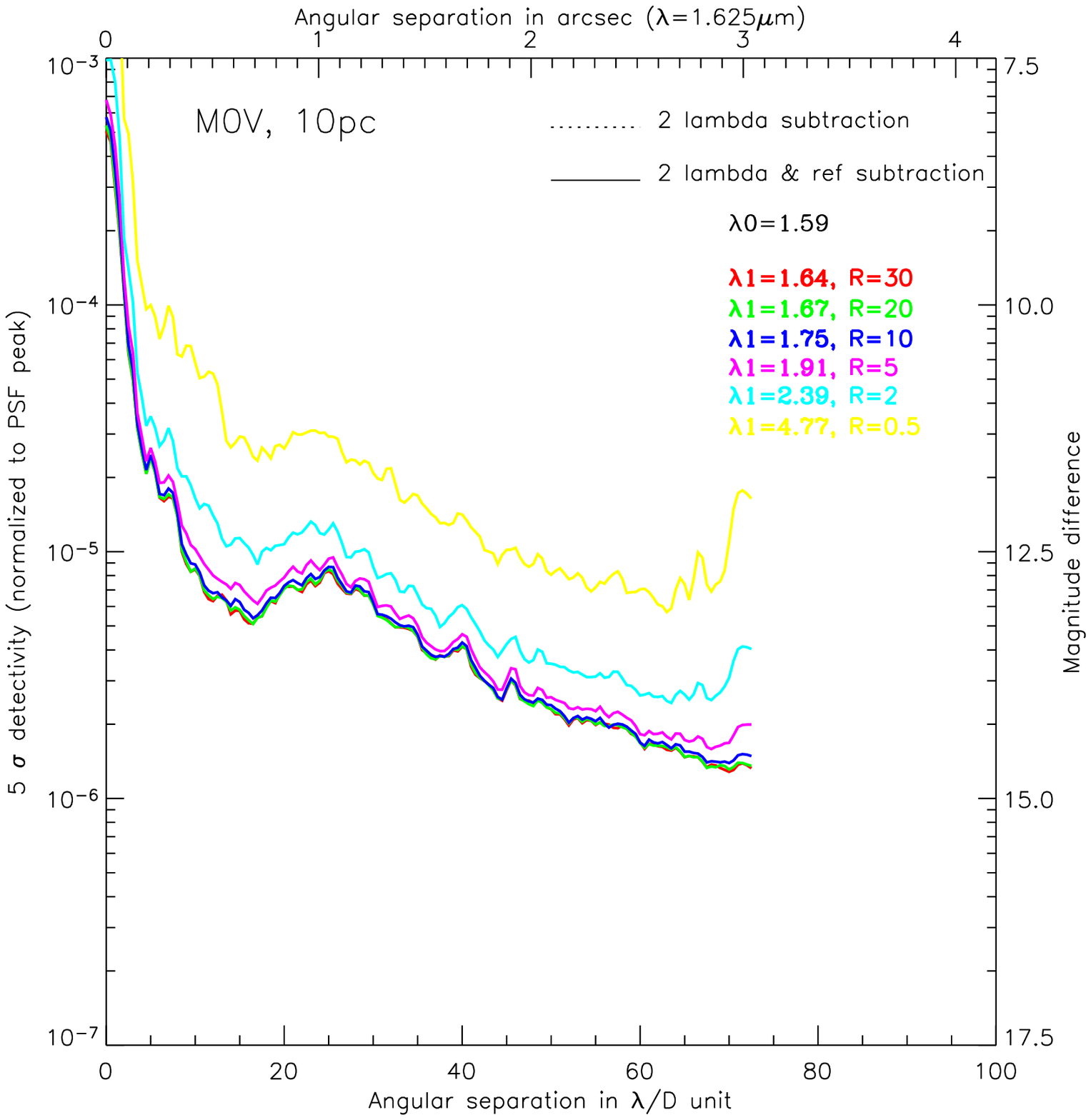}
   \caption{Identical to Fig. \ref{fig:sensitivity1} for the case of differential aberrations (left) and spectral separation between filters (right). }  
    \label{fig:sensitivity2}
\end{figure}

Prior to performance evaluation we carried out a sensitivity analysis of the parameters modeled in the simulation and draw the error budget. Table \ref{tab:params} shows the error budget where each parameters were constrained by simulations while keeping in mind the issue of feasibility. Figures \ref{fig:sensitivity1} and \ref{fig:sensitivity2} give some examples of sensitivity analysis. 

Given those values, we calculated the detectivity for several test cases. Our set of 240 cases covers various observing conditions (turbulence conditions), various science targets (stellar types and ages, planets from 1 to 30 Jupiter masses), and various observing modes (5 filter pairs ranging from 1.08$\mu m$ to 2.25$\mu m$. Planet spectra are taken from Allard et al. (2001) models. The signal to noise ratio is calculated on radial profiles (azimuthally averaged) obtained on the coronagraphic image, on the single subtraction and on the double subtraction. 
A complete set of results makes possible an extensive discussion on the science case of the instrument, and the optimal observing strategy (\cite{moutou05}) . For and old M star (1Gyr) in the solar vicinity a giant planet of about 7 times the mass of Jupiter can be detected. The situation is more favorable for young stars for which 1 M$_J$ planets are detectable in orbit as close as 0.2" (8AU at 40pc) since they are relatively bright and feature strong methane absorption. Regarding nearby old stars (3pc), the detection is also improved to 1 M$_J$ for irradiated planets (separation = 1 AU) according to the model of Sudarsky et al. (2003).

\begin{table}
\centering
\begin{tabular}{ll}\hline\hline
parameters									& 	values			\\ \hline
pupil alignment error in translation 					& 	$<$ 0.2\%			\\
pupil alignment error in rotation 					&	$<$ 0.1$^{\circ}$	\\
image alignment and stability on coronagraph 			&	$<$ 0.5 mas		\\
image defocus on coronagraph					&	$<$ 4 nm rms		\\
coronagraphic phase mask chromatism 				&	$<$ 10$^{-2}$ rd	\\
Lyot stop alignment								&	$<$ 0.2\%			\\
wavefront errors non common to simultaneous images	&	$<$ 10 nm rms		\\
total wavefront errors downstream Lyot stop			&	$<$ 90 nm rms		\\
$\lambda/\Delta\lambda$	between 2 spectral channels	&    	$>$ 10			\\
Flat Field residual errors							&	$<$ 10$^{-3}$		\\ \hline
\end{tabular}
\caption{System defects included in the VLT-PF end-to-end simulations.}
\label{tab:params}
\end{table}

\section{The particular case of Extremely Large Telescopes: terrestrial planets}
Terrestrial planets have became the top priority targets of Extremely Large Telescopes. The feasibility of this program is overwhelming the single astrophysical interest and therefore needs to be evaluated thoroughly. The star/planet contrast is about 10$^{-10}$ at a projected separation of 0.1" for a system placed at 10pc. Therefore, compared to the performance delivered by VLT Planet Finder a gain of 4 orders of magnitude is needed. For the same reason as above a differential technique is needed to improve the detectability.

\begin{figure}[t]
   \centering
      \includegraphics[width=6.5cm]{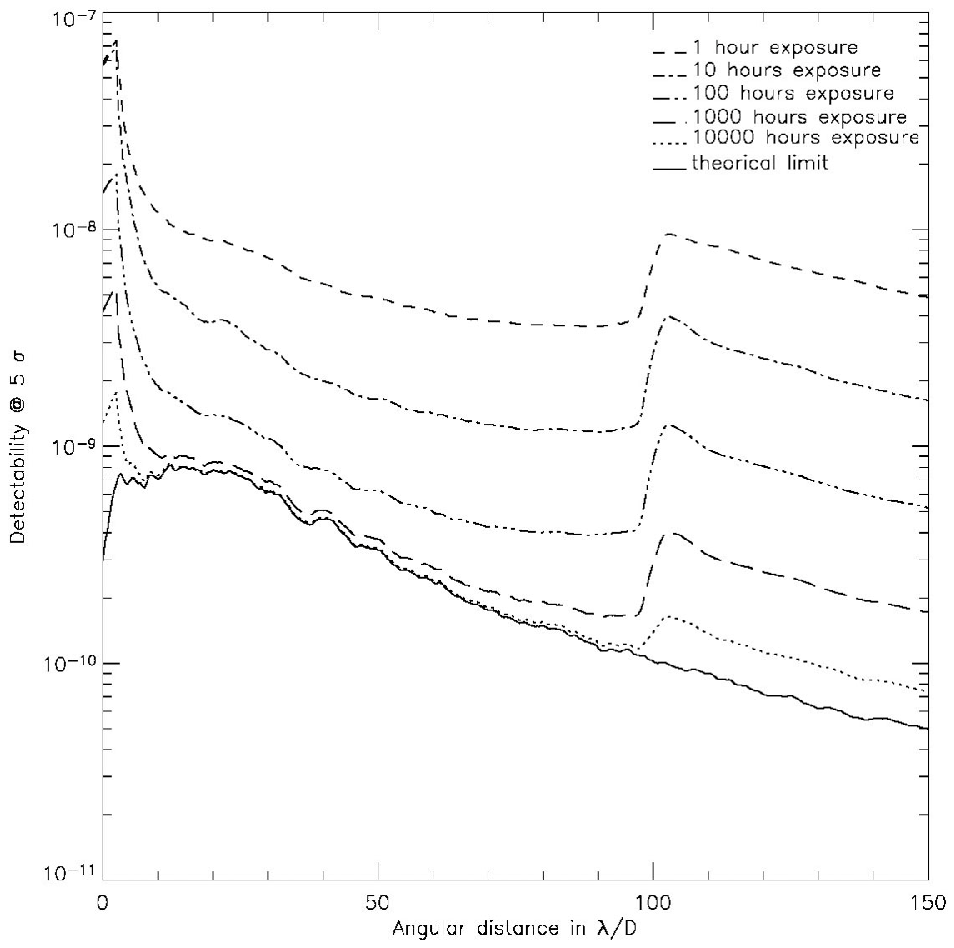}
      \includegraphics[width=6.5cm]{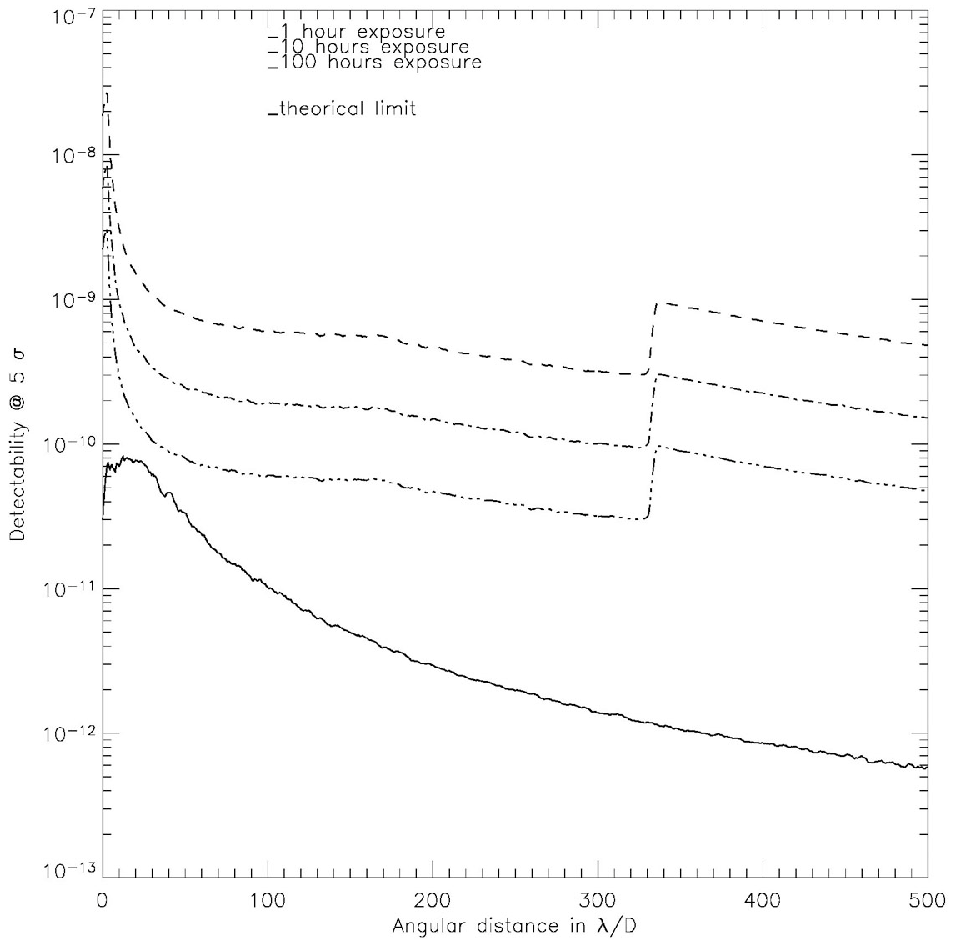}
   \caption{Results of simulation obtained by Cavarroc et al. (2005) using a  simple analytical model of differential imager in ELTs. The left plot stands for a 30m telescope and differential aberrations of 0.1nm rms and the right plots is for a 100m telescope having reduced the differential aberrations to 0.01nm rms. In both cases the common aberrations are set to 20nm rms.}
   \label{fig:elts}
\end{figure}

Our team developed a simple analytical model which is including the atmospheric phase residual left uncorrected by the AO system, the static common aberrations in the instrument and the differential (non common) aberrations  (\cite{cavarroc05}). 
For sake of generality we did not assume any particular type of differential technique. So we just considered two images recorded simultaneously one showing the planet and the other not. 
It is shown that the fundamental limit (independently of the integration time) is given by the amount of static aberrations and that the effects of common and differential optics are correlated. The amplitudes of aberrations is extremely challenging to allow planet detection. For a 100m telescope common and differential optics must be controlled respectively at 20nm rms and 0.01nm rms to achieve a planet/star contrast of 10$^{-10}$.
Moreover, our assumptions on the AO system performance lead to much longer integration time (a few hundreds of hours for a 100m telescope) than what is previously announced in the literature. It is obvious that simulated performance of ELTs cannot be realistic if static aberrations are neglected. The other important result is that if we manage to lower the static aberrations to the aforesaid level we still need a telescope diameter of 100m to achieve the detection. 
Two results of simulation are shown in Fig. \ref{fig:elts} for respectively a 30m telescope with differential aberrations of 0.1nm rms and a 100m telescope with differential aberrations 0.01nm rms.

\section{Conclusion}
Differential techniques are becoming available on current AO systems and are providing first results on young star associations. In the near future, extreme AO systems on 8m ground-based telescopes will be equipped with spectral differential imaging units. Extrasolar Giant Planets are expected to be discovered and characterized using such instruments since the phase aberrations requirements for such programs are within the current capabilities of differential techniques and providing the alignments and pointing inside the coronagraph are accurately controlled. 
For the longer term, ELTs will have the very challenging objective to image terrestrial planets from the ground but the performance of differential techniques will have to be pushed at a limit which is several orders of magnitude above the current state of the art. More developments and modeling will be mandatory in this area.


\end{document}